\documentclass[hyper]{JHEP}
\pdfoutput=1
\usepackage{fixltx2e}
\usepackage{float}
\usepackage{graphicx,amssymb,bm,latexsym,amsmath}
\usepackage{epstopdf}
\usepackage{caption}
\usepackage{subcaption}

\title{A note on oscillating strings in $AdS_3 \times S^3$ with mixed three-form fluxes}
\author{Aritra Banerjee, Kamal L. Panigrahi and Manoranjan Samal\\
Department of Physics,\\Indian Institute of Technology Kharagpur,\\
Kharagpur-721 302, India\\
Email: \email{aritra,panigrahi,manoranjan@phy.iitkgp.ernet.in}}

\abstract{We present a detailed study of the pulsating string solutions
in  $AdS_3 \times S^3$ supported  by both RR and NS-NS 
fluxes. This background has recently been proved to be integrable. We find
the dispersion relation between the energy,  oscillation number and 
other conserved charges when the NS-NS flux turned on is small. We further
discuss the fate of the string  solutions in pure RR and NS-NS cases.}


\begin{document}

\section{Introduction}
The AdS/CFT duality between the $\mathcal{N}= 4$ Supersymmetric Yang-Mills (SYM)
theory in four dimensions and type IIB superstring in the
compactified $AdS_5$ space \cite{Maldacena:1997re} has been the playground 
for research for more than fifteen years. Although finding exact matching
between string states and dual operators from both sides is a highly
non-trivial problem, studying classical string solutions in various $AdS\times S$
backgrounds have played an important role along that direction. The anomalous
dimensions of particular field theory operators having large charges can be
obtained in the string side simply by looking at the dispersion relation among charges
of classical strings. The fact that the dilatation operator 
of the  $\mathcal{N}= 4$ SYM theory in one loop can be written as 
the hamiltonian of the integrable Heisenberg $XXX_{1/2}$ spin chain 
\cite{Minahan:2002ve} has been proved to be a key concept in connecting
integrability, spin chains and string theory in the context of 
classical solutions. This observation helped cast the strong coupling
problem of the SYM theory into solving the algebraic bethe ansatz equations
for the quantum spin chain \cite{Beisert:2003yb}. Also, in the string
theory side, the integrability of the $AdS_5\times S^5$ string sigma model
 based on the supercoset $\frac{PSU(2,2|4)}{SO(4,1)\times SO(5)}$ 
 has been established as the equations of motion for the superstring 
 from this sigma model can be
recast in the zero curvature form \cite{Bena:2003wd}. This ensures the 
existence of an infinite number of conserved
quantities associated to string motion in this background.
To speak precisely, the integrability associated with both sides 
of the duality has improved the
understanding of the equivalence between the Bethe equation for
the spin chain and the corresponding realization of
worldsheet symmetries of the classical $AdS_5 \times S^5$ string sigma
model \cite{Kazakov:2004qf,Zarembo:2004hp}. The
relevant Bethe equations are based on the knowledge of the
S-matrix which focusses on the scattering of world-sheet excitations
of the gauge-fixed string sigma model, or equivalently, the excitations of a
certain spin chain in the dual gauge theory \cite{
Beisert:2004hm,Arutyunov:2004vx,
Staudacher:2004tk,Beisert:2005fw,Beisert:2005tm}.

Over the years, a lot of studies have been done on the semiclassical
rotating string solutions arising from the $AdS_5 \times S^5$ string sigma
model. This includes well known solutions like giant magnons \cite{Hofman:2006xt},
spiky strings \cite{Kruczenski:2004wg} and most importantly folded 
spinning string solutions \cite{Gubser:2002tv}. However 
unlike their rotating counterparts, the pulsating string solutions
\cite{Minahan:2002rc} are less explored even though the 
pulsating-rotating solutions offer better stability than 
the pure rotating ones  \cite{Khan:2005fc}. These solutions 
are time-dependent as opposed to the usual rigidly rotating
string solutions. They are expected to be dual to highly excited
states in terms of operators. For example the most general
pulsating string in $S^5$ charged under the isometry
group $SO(6)$ will have a dual operator of the form
$\text{Tr} \left(X^{J_1}Y^{J_2}Z^{J_3}\right)$.
Here the $X$, $Y$ and $Z$ are the chiral scalars and $J_{i}$'s are 
the R-charges from the SYM theory.  Pulsating string solutions 
have been thoroughly  generalized in \cite{Khan:2003sm,Engquist:2003rn,
Arutyunov:2003za,Dimov:2004xi,Smedback:1998yn,Kruczenski:2004cn},
and have been studied in a number of backgrounds 
having varying degrees of supersymmetry \cite{Chen:2008qq,Dimov:2009rd,
Bobev:2004id,Arnaudov:2010by}. Simultaneously rotating
and oscillating strings were presented in \cite{Park:2005kt}, and the
generalisation of this was done with extra angular momenta in 
\cite{Pradhan:2013sja}. The one loop  corrections to pulsating string spectrum 
in $AdS$ have been discussed in \cite{Beccaria:2010zn}.  Recently these 
type of strings have been used to probe integrable deformations of $AdS$
string models like the Lunin-Maldacena background \cite{Giardino:2011jy} and the
one-parameter deformed $AdS$ backgrounds \cite{Banerjee:2014bca,Panigrahi:2014sia}.
Also such pulsating solutions in non-local geometries have been looked at in \cite{Banerjee:2014rza}.

The other famous example of a holographic dual pair is that of string
propagation in $AdS^3\times S^3 \times T^4$ background and the 
$\mathcal{N}$= (4,4) superconformal field theory coming from the
$D1-D5$ brane system. This duality has also been well explored from both sides
and various semiclassical solutions in this background
\cite{Maldacena:2000hw,misc6,David:2008yk,David:2010yg,Abbott:2012dd,Beccaria:2012kb,
Beccaria:2012pm,Abbott:2013ixa,Rughoonauth:2012qd,Cardona:2014gqa} has been studied
in the context of integrability \footnote{For a review of integrability in
$AdS_3/CFT_2$ one can see \cite{Sfondrini:2014via} and references therein.}
. String theory in this background supported by NS-NS type flux can 
be described in terms of a $SL(2,\mathbb{R})$ WZW model. It has been
 suggested recently that in  $AdS^3\times S^3$ background supported by
both NS-NS and RR type fluxes ( $H_3= dB_2$ and $F_3= dC_2$ ), the string theory is integrable
\cite{Cagnazzo:2012se, Wulff:2014kja}. The 
integrable structure of this theory, including the S-matrix,
has been discussed at length in \cite{Hoare:2013pma,Hoare:2013ida,
Bianchi:2014rfa,Hoare:2013lja,Babichenko:2014yaa,Borsato:2014hja,Lloyd:2014bsa}, and
various semiclassical string solutions have been studied in 
\cite{Hoare:2013lja,David:2014qta,Ahn:2014tua,misc9,Hernandez:2014eta,
Hernandez:2015nba}. The NS-NS flux in this background depends on a 
parameter $q$ with $0 \leq q \leq 1$, while 
the RR flux will be dependent on a parameter $\hat{q}=\sqrt{1-q^2}$.
This model then interpolates between a pure RR background, for which
one can find the spectrum using the integrability based approaches,
and the pure NS-NS case where the WZW approach suffices. But it has 
been clear that for a intermediate value of $q$, none of the approaches
will be suitable enough. Classical string solutions in this mixed flux 
background might be helpful to bridge the two models together.

In the present note, we address the question of finding pulsating 
string solutions in this `mixed-flux' background. We solve the 
 F-string equations of motion with NS-NS flux and find the dispersion 
relations among various conserved quantities perturbatively upto the 
$\mathcal{O}(q^2)$ , provided the flux turned on is small.
To find the quantized spectrum of the string, we will use a 
Bohr-Sommerfeld quantization. Since the
motion of the string is (quasi)periodic, we can use the oscillation
number $N= \oint p~dq$ and the Noether charges to characterize its
dynamics. The oscillation number is an adiabtic invariant quantity and we 
find that in terms of elliptic functions,  which leads to the energy
of the string. We will see that when the NS-NS flux is switched off, our
results  match exactly that of already known ones in the literature. We 
also discuss the dynamics of such strings in pure NS-NS background in 
both small and large energy regimes.

The rest of the note is organized as follows. In section 2, we will 
discuss about the strings pulsating near the centre of the sphere on 
the $\mathbb{R}\times  S^3$ with mixed flux. In section 3, we will talk
about this class of strings in $AdS_3$ with a background mixed flux. We
will also sketch a generalisation of the solution with an angular momentum
in $S^3$ and comment on the effect of its inclusion. In section
4 we conclude and present our outlook.

\section{Pulsating string in ~ $ \mathbb{R}\times  S^3$ with mixed flux}
In this section we will study the semiclassical quantization of a closed string 
which undergoes pulsating motion in the $ \mathbb{R}\times  S^3$ subsector of the total
geometry. Let us start with the metric and the NS-NS flux
\footnote{There is a gauge freedom in the choice of the two-form B-field
for the Wess-Zumino term since the supergravity equations of motion depend on the three-form field strength 
instead. In the presence of a boundary one can incorporate a boundary  
term to parameterise
this ambiguity. For details one can look at \cite{Hoare:2013lja} where the 2-form field is written as 
$-\frac{q}{2}(\cos 2\theta +c)$, with $c$ as the ambiguity term. This $c$ can be fixed via natural physical
requirements.} supporting the 
background,
\begin{eqnarray}
ds^2 &=& -dt^2+ d \theta ^2 + \sin ^2 \theta d\phi _ 1 ^2 + \cos ^2 \theta  d\phi_ 2 ^2 \,\nonumber\\
b_{ \phi_1 \phi_2} &=& -q \cos^2 \theta   \label{metric1}
\end{eqnarray}
Since we are interested in F-string motion, the RR flux will not be relevant for us. 
In what follows throughout the paper, we will take the value of $q$ to be small and use 
it as a perturbation parameter as in \cite{David:2014qta}. Now to study string solutions
in this background, we use the polyakov action 
\begin{equation}
S=-\frac{\sqrt{\lambda}}{4\pi}\int d\sigma d\tau
[\sqrt{-\gamma}\gamma^{\alpha \beta}g_{MN}\partial_{\alpha} X^M
\partial_{\beta}X^N - \epsilon^{\alpha \beta}\partial_{\alpha} X^M
\partial_{\beta}X^N b_{MN}] \ ,
\end{equation}
where  $\sqrt{\lambda}$ is the 't Hooft coupling,
$\gamma^{\alpha \beta}$ is the worldsheet metric and $\epsilon^{\alpha
\beta}$ is the antisymmetric tensor defined as $\epsilon^{\tau
\sigma}=-\epsilon^{\sigma \tau}=1$.
Variation of the action with respect to
$X^M$ gives us the following equation of motion
\begin{eqnarray}
2\partial_{\alpha}(\eta^{\alpha \beta} \partial_{\beta}X^Ng_{KN})
&-& \eta^{\alpha \beta} \partial_{\alpha} X^M \partial_{\beta}
X^N\partial_K g_{MN} - 2\partial_{\alpha}(\epsilon^{\alpha \beta}
\partial_{\beta}X^N b_{KN}) \nonumber \\ &+& \epsilon ^{\alpha \beta}
\partial_{\alpha} X^M \partial_{\beta} X^N\partial_K b_{MN}=0 \ ,
\end{eqnarray}
and variation with respect to the metric gives the two Virasoro
constraints,
\begin{eqnarray}
g_{MN}(\partial_{\tau}X^M \partial_{\tau}X^N +
\partial_{\sigma}X^M \partial_{\sigma}X^N)&=&0 \ ,\label{v1} \\ 
g_{MN}(\partial_{\tau}X^M \partial_{\sigma}X^N)&=&0 \label{v2}\ .
\end{eqnarray}
We use the conformal gauge (i.e.
$\sqrt{-\gamma}\gamma^{\alpha \beta}=\eta^{\alpha \beta}$) with
$\eta^{\tau \tau}=-1$, $\eta^{\sigma \sigma}=1$ and $\eta^{\tau
\sigma}=\eta^{\sigma \tau}=0$) to solve the equations of motion.
Let us propose an ansatz for studying the pulsating string 
in the form,
\begin{equation}
 t=t(\tau), ~ ~ \theta = \theta (\tau) ,~ ~ \phi_1 = m \sigma, ~ ~ \phi_2 = \phi_2(\tau).
\end{equation}
With this embedding, the equations of motion of $ \theta $  and $ \phi _2 $ give
\begin{eqnarray}
\ddot{\theta} &=& - \cos \theta \sin \theta (m^2+ \dot{\phi_2}^2+ 2 m q \dot{\phi_2}),\nonumber\\
\ddot{\phi_2} &=& 2 \tan \theta \dot{\theta }(m q + \dot{\phi_2}). \label{EOM}
\end{eqnarray} 
While the $t$ equation is satisfied simply by
\begin{equation}
 t= \kappa \tau ,
\end{equation}
with $\kappa$ being a constant.
One can check that the virasoro constraint (\ref{v1}) is given by
\begin{eqnarray}
\dot{\theta}^2 = \dot{t} ^2 - m^2 \sin^2 \theta - \cos^2 \theta \dot{\phi _2}^2, \label{vira1}
\end{eqnarray}
while the (\ref{v2}) is trivially satisfied.
To check the consistency of the equation of motion with the virasoro, we can
integrate (\ref{EOM}) to get 
\begin{eqnarray}
 \dot{\theta}^2 &=& - m^2(1-q^2)\sin^2\theta- \frac{C_1^2}{\cos^2\theta}+ C_2,\nonumber\\
 \dot{\phi_2} &=& \frac{C_1}{\cos^2\theta} - m q , \label{eoms1}
\end{eqnarray}
where $C_1$ and $C_2$ are integration constants. It can be easily verified that the
above equations are consistent with (\ref{vira1}) with the choice 
\begin{equation}
 C_2 = \kappa^2 - m^2 q^2 + 2 m q C_1.
\end{equation}
Now, looking at the isometries of the background (\ref{metric1})
we  find the conserved Noether charges are the energy and angular momentum of the
string.
\begin{eqnarray}
 E = \frac{\sqrt{\lambda}}{2\pi}
 \int\dot{t} d\sigma, ~~ ~ J= \frac{\sqrt{\lambda}}{2\pi}
 \int (\cos ^2 \theta  \dot{\phi _2}+  q \cos ^2 \theta \phi_1 ')d\sigma,
\end{eqnarray}
The above can be rewritten as 
\begin{eqnarray}
\mathcal{E} = \dot{t}, ~~ ~ \mathcal{J}= \cos ^2 \theta  \dot{\phi _2}+ m q \cos ^2 \theta,
\end{eqnarray}
since the integrands are not functions of $\sigma$ we can perform the integration accordingly.
Also, we have used $\mathcal{E}= \frac{E}{\sqrt{\lambda}}$ and
$\mathcal{J}= \frac{J}{\sqrt{\lambda}}$ as the `semiclassical' values of the charges. It is
worth noting that due to the gauge freedom in choosing the B-field one could add a constant
term to the definition of the 2-form in  (\ref{metric1}). This would have made the charge 
$\mathcal{J}$ ambiguous too with extra terms proportional to $q$ appearing in the definition.
We can avoid this by choosing the boundary term accordingly so that the field exactly is 
given by  (\ref{metric1}).

Using the above we can write the Virasoro constraint in the following suggestive form,
\begin{eqnarray}
\dot{\theta}^2 = \mathcal{E}^2 - m^2\sin^2\theta  - \frac{\mathcal{J}^2}{\cos^2 \theta}
- m^2 q ^2 \cos ^2 \theta + 2 m \mathcal{J} q.
\end{eqnarray}
It can be seen that as $\theta$ varies between $0$ to $\frac{\pi}{2}$, the $\dot{\theta}^2$ varies between
 $\mathcal{E}^2 - (\mathcal{J}-m q)^2$ to infinity. This looks like the equation of motion of a particle
moving in an effective potential $V(\theta)$, where $\theta$ rotates between a minimal and a maximal
value. Note that the $\theta$ equation can be written in the form,
\begin{equation}
 \ddot{x} = m^2(1- q^2)\left[-(R_+ + R_-)x + 2 x^3 \right],
\end{equation}
where $x = \sin\theta$. The $R_\pm$ are roots of the polynomial 
\begin{equation}
 f(z) = (\mathcal{E}^2 - m^2 q^2+ 2m q \mathcal{J})(1-z) - m^2 z(1 - z)(1-q^2) - \mathcal{J}^2,
\end{equation}
with $z= x^2$. This can be written as
\begin{equation}
 \dot{x}^2=m^2(1-q^2)(x^2-R_-)(x^2- R_+ ), \label{jacobi}
\end{equation}
With arbitrary coefficients, the above equation of motion would represent a undamped Duffing
oscillator. But with proper scaling of the variables, we can write the solution of (\ref{jacobi}) in terms of
standard Jacobi elliptic function\footnote{In the notation we follow $\text{sn}(z\arrowvert m)$ is the
solution of $w'(z)^2 = (1-w^2(z))(1-m w^2(z))$.} provided the initial condition $x(0)=0$: 
\begin{equation}
 \sin\theta(\tau) = {\sqrt{R_-}}~\text{sn}\left(m\sqrt{1-q^2}\sqrt{R_+}\tau \arrowvert \frac{R_-}{R_+}\right).
\end{equation}
Using the property of Jacobi functions that $\text{sn}(z\arrowvert m)= \text{sn}(z+ 4\mathbb{K}(m)\arrowvert m)$
and as usual taking only the real period, we can find that the condition for time-periodic solution for $\theta$
is,
\begin{equation}
 \frac{R_-}{R_+}<1.
\end{equation}
This translates to the following inequality 
\begin{equation}
 (\mathcal{E}^2-m^2)^2+ 4\mathcal{J}^2 m^2 + 4\mathcal{J}m q  (\mathcal{E}^2-m^2)>0, \label{con2}
 \end{equation}
which gives a constraint on the conserved charges so that the string has a pulsating motion.
Since the above quantity is the discriminant of $R_\pm$, this guarantees that the roots are 
real and not equal. It is also important to mention that for a periodic solution both  $R_\pm$
have to be positive, which leads to the following,
\begin{equation}
 \mathcal{E}^2 - (\mathcal{J}-m q)^2 > 0. \label{condi1}
\end{equation}
This is in tune with our observation earlier about the limits of oscillation in the $\dot\theta$ equation.

We can also find the dynamics of the string along the $\phi_2$ direction by integrating 
$\frac{d\phi_2}{d\theta}$ from (\ref{eoms1}) which has the form
\begin{equation}
 \frac{d\phi_2}{d\theta}= \frac{\mathcal{J}- m q \cos^2\theta}{m\sqrt{1-q^2}\cos\theta 
 \sqrt{(\sin^2\theta-R_-)(\sin^2\theta-R_+)}}.
\end{equation}
This can be integrated to find $\phi_2$ in terms of standard elliptic integrals,
\begin{eqnarray}
 \phi_2(\tau)&=& \frac{1}{m\sqrt{1-q^2}}\Bigg[\frac{\mathcal{J}}{\sqrt{R_+}}\Pi
 \left( R_-,\sin^{-1}(\frac{1}{\sqrt{R_-}}\sin\theta(\tau)),\frac{R_-}{R_+}\right) \nonumber\\
 &-&\frac{mq}{\sqrt{R_+}}\mathbb{F}\left(\sin^{-1}(\frac{1}{\sqrt{R_-}}\sin\theta(\tau))
 ,\frac{R_-}{R_+}\right)\Bigg] \label{phi2}
\end{eqnarray}

\begin{figure}[h]
   \centering
        \begin{subfigure}[b]{0.2\textwidth}
                \includegraphics[width=\textwidth]{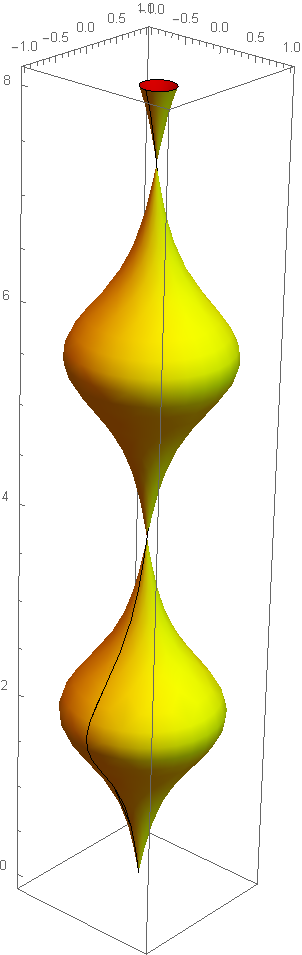}
                \caption{}
                \label{fig:1.1}
        \end{subfigure}%
        ~ 
        \begin{subfigure}[b]{0.2\textwidth}
                \includegraphics[width=\textwidth]{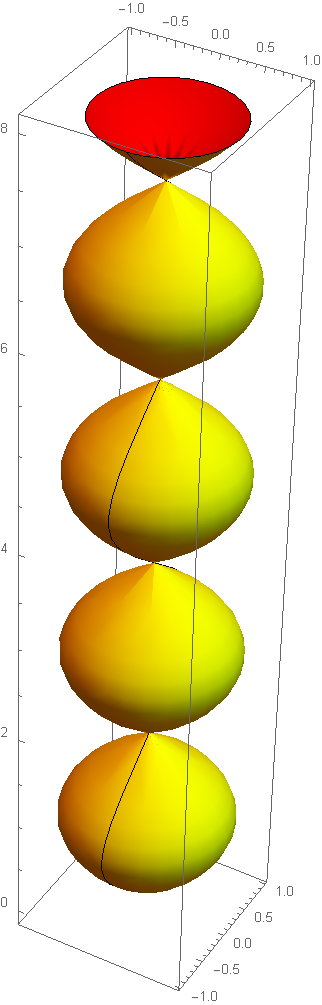}
                \caption{}
                \label{fig:1.2}
        \end{subfigure}
        \caption{Pulsating `short' circular strings in $\mathbb{R}\times S^3$ as introduced in (\ref{parameter1})
        evolving in the $(X^0, X^1, X^2)$ plane. 
        Both the strings have $m=2$, $\mathcal{E}=\mathcal{J}=0.2$ with a) $q=0.25$ and
        b) $q=0.99$. The vertical direction coincides with the worldsheet time $\tau$.}\label{fig:1}
\end{figure}

\begin{figure}[h]
        \centering

                \includegraphics[width=0.6\linewidth]{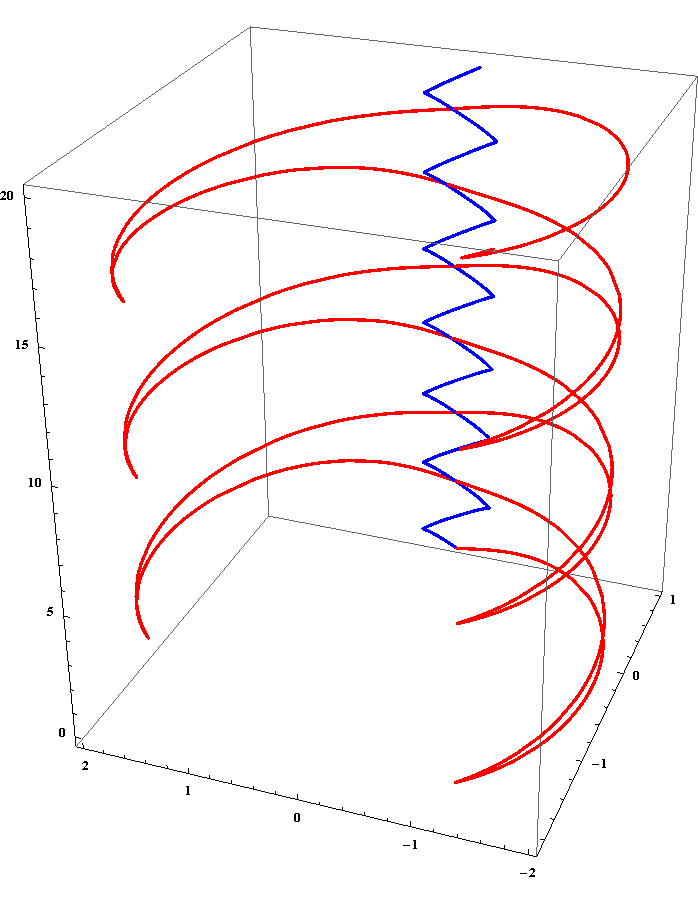}

        \caption{Plot of temporal evolution of $\phi_2(\tau)$ as described by eq (\ref{phi2}) in the 
        $(X^0, X^3, X^4)$ plane. Both the trajectories have $m=2$, $\mathcal{E}=0.2$, $\mathcal{J}=0.5$. Blue one 
        corresponds to
        $q= 0.02$ and Red one corresponds to $q=0.95$. The trajectories zig-zag along as the strings evolve. The 
        vertical axis represents the worldsheet time $\tau$.}\label{fig:33}
\end{figure}

Now since the solutions we are looking for is that of a circular string, we can parameterise 
the worldsheet in a simplified way using the variables
\begin{eqnarray}
 X^0 &=& \tau, \nonumber\\
 X^1 &=& \sin\theta(\tau) \cos (m \sigma), \nonumber\\
 X^2 &=& \sin\theta(\tau) \sin (m \sigma), \nonumber\\ 
 X^3 &=& \cos\theta(\tau)\cos\phi_2(\tau), \nonumber\\
 X^4 &=& \cos\theta(\tau)\sin\phi_2(\tau),~~~~~~~~\sigma\in[0,2\pi]. \label{parameter1}
\end{eqnarray}
These are just some naive representations of the hypersurface traced out by the string
as it moves forward in $\tau$ (in analogy with the flat space pulsating solutions obtained in
\cite{Minahan:2002rc}), the actual worldsheet in a curved background can be very complex.
The string solutions in $(X^0, X^1, X^2)$ plane are plotted in for various values of the
parameters in figure (\ref{fig:1}). The motion along $(X^0, X^3, X^4)$ plane has been visualised
in  figure (\ref{fig:33}) to demonstrate how the $\phi_2$ coordinate evolves with time.

Now we will use Bohr-Sommerfeld like quantization procedure for the pulsating strings in this
background. The oscillation number (the adiabatic invariant associated to $\theta$) can be written using
the canonical momenta conjugate to $\theta$ as follows,
\begin{eqnarray}
N &=& \sqrt{\lambda} \mathcal{N} = \frac{\sqrt{{\lambda}}}
{2\pi} \oint d\theta ~ \Pi _\theta \cr & \cr &=& \frac{\sqrt{\lambda}}{2\pi} \oint d\theta 
\sqrt{\mathcal{E} ^2 - m^2 \sin^2 \theta - \frac{\mathcal{J}^2}{ \cos^2 \theta}
- m^2 q^2 \cos^2\theta + 2 m  \mathcal{J} q}.
\end{eqnarray}
Again, taking  $\sin \theta = x $ we can choose the proper limits and transform the above integral to
\begin{equation}
\mathcal{N} =\frac{2 m\sqrt{1- q^2}}{\pi} \int _0 ^ {\sqrt{R_-}}  \frac{d x}{ (1 - x^2)} 
\sqrt{(x^2 - R_+)(x^2 - R_-)}. \label{N1}
\end{equation}
We can directly compute the integral to find,
\begin{eqnarray}
 \mathcal{N}= \frac{2m\sqrt{1-q^2}}{\pi\sqrt{R_+ - R_-}}\Bigg[(R_+ - R_-)\mathbb{E}\left(\frac{-R_-}{R_+ - R_-}\right)
 &+& (R_- -1)\mathbb{K}\left(\frac{-R_-}{R_+ - R_-}\right) \nonumber\\
 -(R_+ - 1)\Pi\left(\frac{R_-}{R_- -1},\frac{-R_-}{R_+ - R_-}\right)\Bigg].
\end{eqnarray}
Instead of working with this, we can make the expressions a little simpler by
taking the partial derivative of (\ref{N1}) with respect to $m$ leading to 
 \begin{eqnarray} 
\frac{ \partial\mathcal{N}}{\partial m} = I_1  - I_2.
\end{eqnarray}
 The integrals $I_1$ and $I_2$ are given by 
\begin{eqnarray}
 I_1 &=& \frac{2q (\mathcal{J} -m q)}{\pi m \sqrt{1 - q^2}} \int _0 ^ {\sqrt{R_-}}  
 \frac{d x}{ \sqrt{(x^2 - R_+)(x^2 - R_-)}},\nonumber \\
 I_2 &=& \frac{2 \sqrt{1 - q^2}}{\pi} \int _0 ^ {\sqrt{R_-}}  
 \frac{x^2~ d x}{ \sqrt{(x^2 - R_+)(x^2 - R_-)}}.
\end{eqnarray}
The integrals evaluate to complete elliptic functions of the first and second kind, and
gives the expression
\begin{eqnarray}
 \frac{ \partial\mathcal{N}}{\partial m}=\frac{2 \sqrt{R_+}\sqrt{1 - q^2}}{\pi}\left[ \mathbb{E} 
 \left( \frac{R_-}{R_+}\right)- \mathbb{K}\left( \frac{R_-}{R_+} \right ) \right]
 +\frac{2q (\mathcal{J} -m q)}{ \pi m \sqrt{1 - q^2} \sqrt{R_+}} \mathbb{K}\left(\frac{R_-}{R_+}\right). \nonumber\\
\end{eqnarray}
In short string limit, i.e. when both energy and angular momentum of the string are small, we 
can expand the above expression keeping upto $\mathcal{O}(q^2)$ correction terms. After expanding,
we can integrate to get the series for $\mathcal{N}$ as,
\begin{eqnarray}
\mathcal{N}&=&\Bigg[-\frac{m q^2}{2}+\left (-\frac{1}{2 m}+ \frac{3 q^2}{8m}\right )\mathcal{J}^2 
-\frac{3q \mathcal{J}^3}{4 m^2} +
\left(\frac{5}{16 m^3}-\frac{225 q^2}{128 m^3}\right)\mathcal{J}^4
+\mathcal{O}(\mathcal{J}^5)\Bigg] \nonumber \\ &+&\Bigg[\left (\frac{1}{2 m} +\frac{ q^2}{8 m}\right) 
+ \frac{ q \mathcal{J}}{4 m^2}+ \left(-\frac{3}{8 m^3}+ \frac{27 q^2}{64 m^3}\right)\mathcal{J}^2 - 
\frac{45 q \mathcal{J}^3}{32 m^4}   \nonumber\\
&+& \left(\frac{105}{128 m^5}- \frac{2625 q^2}{512 m^6}\right)\mathcal{J}^4 
+\mathcal{O}(\mathcal{J}^5) \Bigg] \mathcal{E}^2 + 
\mathcal{O}(\mathcal{E} ^4).
\end{eqnarray}
We can now invert the series to find the expression for the string energy in terms of the
conserved quantities
\begin{eqnarray}
\mathcal{E}=\sqrt{2 m \mathcal{L}} ~ \mathcal{A}_1
(\mathcal{J})\Big[ 1 + \mathcal{B}_1(\mathcal{J})\mathcal{L} + \mathcal{O}(\mathcal{L}^2) \Big ], \label{energy1}
\end{eqnarray}
where 
\begin{eqnarray}
\mathcal{L}&=& \mathcal{N}+\frac{m q^2}{2}+\left (\frac{1}{2 m}- \frac{3 q^2}{8m}\right )\mathcal{J}^2 
+\frac{3q \mathcal{J}^3}{4 m^2}- \left(\frac{5}{16 m^3}-\frac{225 q^2}{128 m^3}\right)\mathcal{J}^4
+\mathcal{O}(\mathcal{J}^5),\nonumber\\
\mathcal{A}_1(\mathcal{J})&=& \Bigg[\left(1+\frac{q^2}{4}\right)+\frac{q\mathcal{J}}{2m}-\left(\frac{3}{4m^2}
-\frac{27q^2}{32m^2}\right)\mathcal{J}^2 \nonumber\\
&-&\frac{45q\mathcal{J}^3}{16m^3} 
+ \left(\frac{105}{64m^4}-\frac{2625q^2}{256m^4}
\right)\mathcal{J}^4 +\mathcal{O}(\mathcal{J}^5)\Bigg]^{-1/2}, \nonumber \\
\mathcal{B}_1(\mathcal{J})&=& \left(-\frac{1}{8m}+\frac{q^2}{64m}\right)-\frac{5q\mathcal{J}}{32m^2}+
\left(\frac{33}{64m^3}-\frac{195 q^2}{256 m^3}\right)\mathcal{J}^2 \nonumber\\
&+& \frac{327q\mathcal{J}^3}{128 m^4} - \left(\frac{933}{512 m^5}-\frac{102963 q^2}{8192 m^5}\right)\mathcal{J}^4
+\mathcal{O}(\mathcal{J}^5).\nonumber
\end{eqnarray}
As expected this expression is  perfectly regular. 
One can explicitly check that for $q=0$, i.e. without any flux the leading orders of string energy 
(\ref{energy1}) reduces to that presented in \cite{Pradhan:2013sja}. Also with $\mathcal{J}=0$ 
the expression reduces to the result obtained in \cite{Beccaria:2010zn} for pulsating strings
in $\mathbb{R}\times S^2$, which reads
\begin{equation}
 \mathcal{E}_{\mathbb{R}\times S^2}= \sqrt{2 m \mathcal{N}}\left[1- \frac{\mathcal{N}}{8m}+\mathcal{O}(\mathcal{N}^2)
 \right].
\end{equation}

\subsection*{The case of $q = 1$: Pure NS-NS flux}
\begin{figure}[h]
        \centering

                \includegraphics[width=0.2\linewidth]{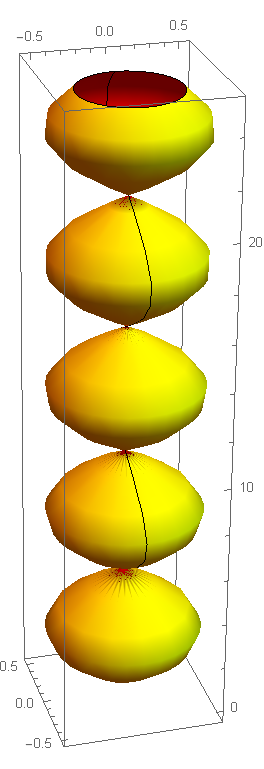}

        \caption{String pulsating in  $\mathbb{R}\times S^3$ supported by pure
        NS-NS flux as described by eq (\ref{pure1}). Here $\mathcal{E}=0.6$, $\mathcal{J}=0.5$ and $m=1$. The 
        vertical axis represents $\tau$.}\label{fig:2}
\end{figure}

As is obvious, the description of pulsating strings we just presented will not be applicable to the
case where $q=1$, i.e. there is only NS-NS flux present. We have to start from the level of equations
of motion and modify the dynamics accordingly.\\
A simple calculation gives the equation of motion for $\theta$ for this case,
\begin{eqnarray}
\dot{\theta}^2 = \mathcal{E}^2 - m^2  - \frac{\mathcal{J}^2}{\cos^2 \theta}
 + 2 m \mathcal{J}.
\end{eqnarray}
With $\sin\theta = x$, the above equation can be transformed into
\begin{equation}
 \dot{x}^2 = p^2(r^2- x^2),
\end{equation}
where $p = \sqrt{\mathcal{E}^2 - m^2+ 2 m \mathcal{J}} $ and $r = \sqrt{\frac{\mathcal{E}^2 - (\mathcal{J}-m )^2}{p^2}}$.
The solution of the equation can be written as 
\begin{equation}
 x(\tau) = \pm r\sin(p\tau) \label{pure1}
\end{equation}
This is a simple pulsating string solution in the regime where $p$ is
real which cannot be true for a short string (small energy) with large $m$. So, a short string pulsating in
$ \mathbb{R}\times  S^3$ must have small $(\mathcal{O}(1))$ number of windings to keep its motion intact.
Such a solution has been visualised in figure (\ref{fig:2}). It is worth noting that the reality condition
of $p$ follows directly from (\ref{con2}) if we simply put $q=1$ and this in turn guarantees the reality of $r$ 
since the numerator under root is nothing but (\ref{condi1}) with $q=1$.
Now, the oscillation number for such a solution can be written as
\begin{equation}
 \mathcal{N} = \frac{2}{\pi}\int^{r}_0 \frac{d x}{1-x^2}\sqrt{p^2(1-x^2)-\mathcal{J}^2}.
\end{equation}
With a transformation of the variable $x \rightarrow r x$, we can write this as
\begin{equation}
 \mathcal{N}= \frac{2}{\pi}r^2 p \int_{0}^{1}\frac{d x}{1-r^2 x^2}\sqrt{1-x^2} 
 = \left(p-\mathcal{J}\right).
\end{equation}
This leads the a simplified expression for string energy
\begin{equation}
 \mathcal{E}= \sqrt{m^2 - 2 m \mathcal{J}+ \left(\mathcal{N}+\mathcal{J}\right)^2}.
\end{equation}
Expanding this in the short string limit, we get,
\begin{equation}
 \mathcal{E}= \left(m - \mathcal{J}\right)+
 \left(\frac{ \mathcal{J}}{m} + \frac{\mathcal{J}^2}{m^2}+ \mathcal{O}(\mathcal{J}^3)\right)\mathcal{N}
 +\left(\frac{1}{2m} + \frac{\mathcal{J}}{2 m^2}+ \mathcal{O}(\mathcal{J}^3)\right)\mathcal{N}^2
 +\mathcal{O}(\mathcal{N}^3).
\end{equation}
\section{Pulsating string in  $AdS_3$ with mixed flux }
\begin{figure}[h]
      \centering
        \begin{subfigure}[b]{0.22\textwidth}
                \includegraphics[width=\textwidth]{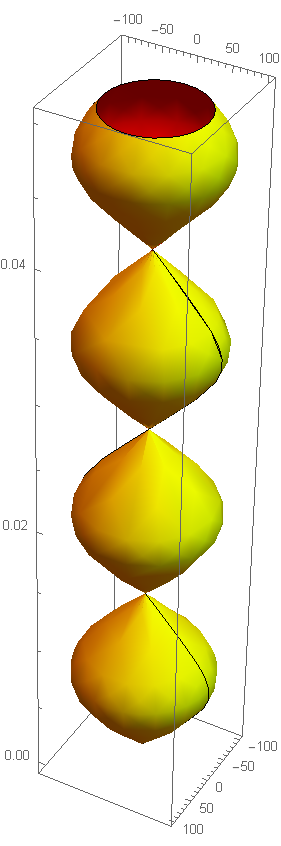}
                \caption{}
                \label{fig:2.3}
        \end{subfigure}%
         \begin{subfigure}[b]{0.205\textwidth}
                \includegraphics[width=\textwidth]{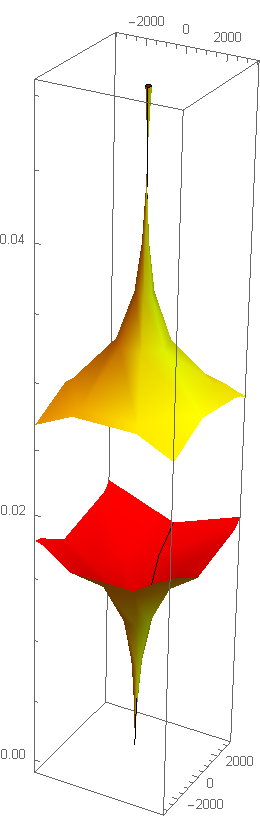}
                \caption{}
                \label{fig:2.4}
        \end{subfigure}
        \caption{Circular `Long' strings in $AdS_3$ plotted according to eq (\ref{AdS_soln}).
        Both of these have $m=2$ and $\mathcal{E}=2\times 10^4$. 
        While a)corresponds to $q=0.1$ and shows perfect pulsating motion, 
        b)corresponds to $q=0.9999$ and does not
        oscillate, giving unphysical structure as the (\ref{cond1}) is alsmost saturated.}\label{fig:3}
\end{figure}

In the current section, we will concentrate on a string which pulsates in $AdS_3$ with mixed flux.
The string contracts to a point in
$\rho= 0$ and then expands up to reach a maximal size $\rho=\rho_{max}$
to contract again and so on.
Later we will also include an angular momentum from the sphere in the solution. 
 We will study the semiclassical quantization of the classes 
of strings which rotates near the centre of $AdS$ and also long objects that go
up to the boundary. One can also do a Hamiltonian analysis of this system by reducing it
to one-dimensional dynamical system. For a simple sketch of this point of view, refer to
the Appendix. Let us start with the metric and NS-NS flux of the relevant background.
\begin{eqnarray}
ds^2 &=& - \cosh^2\rho dt^2 +
d\rho^2 + \sinh^2\rho d\phi^2,\nonumber\\
b_{ t \phi}&=& q \sinh^2 \rho.  \label{metric2}
\end{eqnarray}
We chose the pulsating string ansatz for this configuration as
\begin{eqnarray}
t = t(\tau) ,~~~~~ \rho = \rho(\tau), ~~~~~  \phi &=& m \sigma.
\label{ansatz2}
\end{eqnarray}
 The polyakov action for the given background is given by
\begin{eqnarray}
S = \frac{\sqrt{\lambda}}{4 \pi} \int d\tau d\sigma
\left[- \cosh^2\rho \dot{t}^2 +
\dot\rho^2 - m^2\sinh^2\rho + 2 \dot{t} m q \sinh^2 \rho\right]. \label{action2}
\end{eqnarray}
Equation of motion for $t$ and $\rho$  are given by
\begin{eqnarray}
2\cosh\rho \sinh\rho \dot{\rho}\dot{t}+ \cosh^2\rho\ddot{t}
- 2 m q \cosh\rho \sinh\rho \dot{\rho} &=& 0 ,\nonumber\\
\ddot{\rho}+ \cosh\rho \sinh\rho \left(\dot{t}^2+m^2- 2\dot{t}m q\right) &=& 0 .
\end{eqnarray}
Also from the Virasoro constraints we get
\begin{eqnarray}
m^2\sinh^2\rho - \cosh^2\rho\dot{t}^2+\dot{\rho}^2 = 0. \label{vira2}
\end{eqnarray}
Now, as in the previous section we can integrate the equations of motion to get
\begin{eqnarray}
\dot{t} &=& \frac{C_3}{\cosh^2\rho}+ m q ,\nonumber\\
\dot{\rho}^2 &=& -m^2(1-q^2)\sinh^2\rho + \frac{C_3^2}{\cosh^2\rho}+C_4 , 
\label{eomads}
\end{eqnarray}
where $C_3$ and $C_4$ are integration constants. We can use the above expression
and the (\ref{vira2}) to show that the equations of motion and the Virasoro are completely
self-consistent with the choice,
\begin{equation}
 C_4 = 2 C_3 m q+ m^2 q^2.
\end{equation}
The energy of the oscillating string is given by the conserved charge
\begin{eqnarray}
\mathcal{E} = \cosh^2\rho
 \dot{t} - mq \sinh^2 \rho . \label{25}
\end{eqnarray}
And the canonical momentum conjugate to $\rho$ is
\begin{equation}
\Pi_{\rho} = \dot{\rho} .
\label{26}
\end{equation}
Using the equations (\ref{eomads}) and (\ref{25}), we can write the $\rho$ equation in 
the form
\begin{eqnarray}
\dot{\rho}^2 = \frac{(\mathcal{E}  + m q \sinh^2 \rho)^2}
{\cosh^2 \rho} - m^2\sinh^2\rho  .
\label{27}
\end{eqnarray}
The above may be interpreted as an equation for a particle moving in an
effective potential which grows to infinity at $\rho \rightarrow
\infty$. The coordinate $\rho(\tau)$ thus oscillates between $0$
and a maximal $\rho$ value $(\rho_{max})$. The equation of motion for the 
string can easily be recast using $\sinh\rho = x$ as
\begin{equation}
\dot{x}^2=m^2(1-q^2)(x^2-R_-)(R_+ - x^2),
\end{equation}
where $R_\pm$ are the roots of the polynomial
\begin{equation}
g(z) = (\mathcal{E}- mq)^2 - m^2(1-q^2)z(1+z)+m q (2\mathcal{E} - mq)(1+z).
\end{equation}
The solution can again be written in terms of the Jacobi elliptic function
\footnote{In the notation we follow, $\text{sd}(z\arrowvert m)$ is the solution to
 $w''(z)+ w(z)(2m(1-m)w^2(z)-2m+1) = 0$.}
\begin{equation}
\sinh\rho(\tau) =  \sqrt{\frac{-R_+ R_-}{R_+ - R_-}}~\text{sd}\left(m\sqrt{1-q^2}\sqrt{R_+ - R_-}
 \tau \arrowvert \frac{R_+}{R_+ - R_-}\right). \label{AdS_soln}
\end{equation}
As in the previous section, to have a time-periodic pulsating solution we must have
\begin{equation}
 \frac{R_+}{R_+ - R_-}<1 .
\end{equation}
The above condition translates to the following inequality,
\begin{equation}
 -(m-2 q \mathcal{E})-\sqrt{m^2- 4 m q \mathcal{E}+ 4 \mathcal{E}^2}<0,\label{cond1}
\end{equation}
which gives the constraint on the solution to have a pulsating nature.
Notice that this also takes into account the condition on the roots that $R_-<0$.
Again we can parameterise a circular string motion in terms of,
\begin{eqnarray}
 X^0 &=& \tau, \nonumber\\
 X^1 &=& \sinh\rho(\tau) \cos(m\sigma), \nonumber\\
 X^2 &=& \sinh\rho(\tau) \sin(m\sigma),  ~~~~~~~~\sigma\in[0,2\pi].
\end{eqnarray}
We plot the solution for large value of $\mathcal{E}$ and different values of $q$ to
explore the qualitative behaviour of the string dynamics in figure (\ref{fig:3}). It can be
seen here that with large values of $\mathcal{E}$ and $q\rightarrow 1$, the pulsating motion
of the string is lost, as it progresses towards violating the inequality (\ref{cond1}). To be clear, it can be shown
that in the limit $\mathcal{E}\rightarrow \infty$ and $q\rightarrow 1$, we can write
\begin{equation}
 \frac{R_+}{R_+ - R_-}= 1 + \frac{q-1}{2} + \mathcal{O}\left(\frac{1}{\mathcal{E}}\right),
\end{equation}
which justifies the claim that in this limit the string solutions lose the pulsating motion.
We will talk about it more in a later subsection where we discuss the case with pure NS-NS flux.

We can now define the
oscillation number associated with string motion along $\rho$ direction
as,
\begin{eqnarray}
N &=& \sqrt{\lambda} ~\mathcal{N} =
\frac{\sqrt{\lambda}} {2\pi} \oint d\rho ~ \Pi _\rho \cr &
\cr &=& \frac{2 \sqrt{\lambda}} {\pi} \int_0^{\rho_{max}} d\rho~
\sqrt{\frac{(\mathcal{E} +m q \sinh^2 \rho )^2}{\cosh^2\rho} - m^2\sinh^2\rho}
 ~. \label{29}
\end{eqnarray} 
Changing the  variable to $x=\sinh \rho$  as before we integrate the above to get
\begin{eqnarray}
\mathcal{N}
&=&\frac{2m\sqrt{1 - q^2}} {\pi}\int_0^{\sqrt{R_{+}}}\frac{d
x}{1+x^2}\sqrt{(R_{+}-x^2)(x^2-R_{-})} \\
&=&
\frac{2m\sqrt{1 - q^2}}{\pi}\frac{1}{\sqrt{-R_{-}}}\bigg[R_{-} \mathbb{E}\left(\frac{R_{+}}{R_{-}}\right)
+(1+R_{+})\big[\mathbb{K}\left(\frac{R_{+}}{R_{-}}\right)
-(1+R_{-}) \Pi(-R_{+},\frac{R_{+}}{R_{-}})\big]\bigg]\ . \nonumber
\end{eqnarray}
In the short string limit, the string moves near the centre of $AdS$, where
 $\mathcal{E}$ and $\mathcal{N}$ are small. Remember since the oscillation
 number is counterpart of  the oscillator number $(N_L + N_R)$ in flat space,
 these short strings are not highly excited. In this case we get by expanding in $\mathcal{E}$
 and keeping upto $\mathcal{O}(q^2)$ terms in the coefficients ,
 \begin{eqnarray}
\mathcal{N}= \frac{\mathcal{E}^2}{2m}+ \frac{q \mathcal{E}^3}{2 m^2}+
\frac{5(-1+ 3 q^2)\mathcal{E}^4}{16 m^3}- \frac{21q\mathcal{E}^5}{16 m^4}
+\frac{21(6- 55q^2)\mathcal{E}^6}{256 m^5}+\mathcal{O}(\mathcal{E}^7).
\end{eqnarray}
Inverting the series we get the expression for the string energy,
\begin{eqnarray}
\mathcal{E} =\sqrt{2 m \mathcal{N}} - q \mathcal{N} - 
\frac{5 (-1+ q^2)\mathcal{N}^\frac{3}{2}}{4 \sqrt{2 m}}+ 
\frac{3q\mathcal{N}^2}{2m}+\frac{7(-11+ 24q^2)\mathcal{N}^\frac{5}{2}}{64\sqrt{2}m^\frac{3}{2}}
+\mathcal{O}(\mathcal{N}^3). \nonumber\\
\end{eqnarray}
In the long string limit, the strings are highly excited and can reach the boundary of $AdS$. In this case
$\mathcal{E}$ will be large. For this regime, we can expand the oscillation number in $\mathcal{E}$ and collect
the coefficients order by order.
\begin{eqnarray}
\mathcal{N} = \frac{\mathcal{E}}{2} &+&
\mathcal{C}_1(q)\sqrt{m} \sqrt{\mathcal{E}}-\frac{m q}{2} + \mathcal{C}_2(q)\frac{m^\frac{3}{2}}
{\sqrt{\mathcal{E}}}+ \mathcal{O}(\mathcal{E}^{-\frac{3}{2}}) \nonumber \\
\end{eqnarray}
Where $\mathcal{C}_i(q)$'s have complicated expressions involving elliptic functions of $q$.
Keeping in tune with rest of the paper we can write each of the coefficients as series in powers of $q$,
\begin{eqnarray}
 \mathcal{C}_1(q)&=& -0.38138 + 0.41731 q + 0.14302 q^2+...\nonumber\\
 \mathcal{C}_2(q)&=& 0.069552 + 0.143017 q - 0.13041 q^2+...
\end{eqnarray}
Inverting the series, the long string energy is obtained as,
\begin{eqnarray}
\mathcal{E}= 2 \mathcal{N} + \tilde{\mathcal{C}_1}(q)\sqrt{m \mathcal{N}} +  
\tilde{\mathcal{C}_2}(q)m + \tilde{\mathcal{C}_3}(q)  
\frac{m^{\frac{3}{2}}}{\sqrt{\mathcal{N}}} + \mathcal{O}(\mathcal{N}^{-\frac{3}{2}}). \label{AdSE}
\end{eqnarray}
Again here we can expand and write the coeffiecients as,
\begin{eqnarray}
 \tilde{\mathcal{C}_1}(q)&=&  1.07871 - 1.18034 q - 0.404514 q^2+...\nonumber\\
 \tilde{\mathcal{C}_2}(q)&=& 0.29090 - 1.63662 q + 0.13013 q^2+...\nonumber\\
 \tilde{\mathcal{C}_3}(q)&=&  -0.05914 -0.600694 q +  0.57628q^2+...
\end{eqnarray}
We now have eq (\ref{AdSE}) as the the semiclassical energy expressions for pulsating string configuration
in $AdS_3$ supported by mixed flux. As usual, we have taken the
amount of NS-NS flux turned on to be small and kept upto 
only $\mathcal{O}(q^2)$ terms in the expressions we have considered. To avoid any confusion we must note 
that in the above results
all the coeffiecients have been found exactly and then the values were numerically written down by expanding
them in $q$. For example
 $\tilde{\mathcal{C}_1}(0)=  \frac{4\sqrt{2}}{\pi}[\mathbb{E}(-1)-\mathbb{K}(-1)]\approx1.07871$.
 It also streses on the issue that after
substituting $q = 0$, we can get back  the exact results found 
in \cite{Park:2005kt} and \cite{Beccaria:2010zn}.
\subsection*{Pulsating  string in $(AdS_3 \times S^1)$ with mixed flux }
Similarly as before we would start with the metric of $AdS_3$ with an added
$S^1 \subset S^3$ and the NS-NS flux
\begin{eqnarray}
ds^2 &=& - \cosh^2\rho dt^2 +
d\rho^2 + \sinh^2\rho d\phi^2+ d\psi ^2 \nonumber\\
b_{ t \phi} &=& q \sinh^2 \rho .
\end{eqnarray}
We chose the embedding ansatz for the pulsating string configuration as
\begin{eqnarray}
t = t(\tau) ,~~~~~ \rho = \rho(\tau), ~~~~~  \phi &=& m \sigma, ~~~~~ \psi=\psi(\tau).
\label{21}
\end{eqnarray}
The polyakov action for the given configuration is given by
\begin{eqnarray}
S = \frac{\sqrt{\lambda}}{4 \pi} \int d\tau d\sigma
\left[- \cosh^2\rho \dot{t}^2 +
\dot\rho^2 - m^2\sinh^2\rho + \dot{\psi}^2 + 2 \dot{t} m q \sinh^2 \rho\right]. \label{22}
\end{eqnarray}
The energy and angular momentum  of the oscillating string  are given by
\begin{eqnarray}
\mathcal{E} = \cosh^2\rho
 \dot{t} - mq \sinh^2 \rho,~~~~~~
 \mathcal{J}=\dot{\psi}
\end{eqnarray}
Using the equations of motion and the Virasoro constraints, we can get
\begin{eqnarray}
\dot{\rho}^2 - \frac{(\mathcal{E}  + m q \sinh^2 \rho)^2}
{\cosh^2 \rho} + m^2\sinh^2\rho + \mathcal{J}^2  = 0,
\label{27b}
\end{eqnarray}
which solves the same form of function as in (\ref{AdS_soln}). The qualitative difference
between the solution in (\ref{AdS_soln}) and the one with inclusion of angular momentum
can be discussed accordingly from the solution and the constraints on pulsating solution can
be found. We note here that due to the inclusion of the
extra angular momentum the condition (\ref{cond1}) changes to.
\begin{equation}
 -\mathcal{J}^2-m^2+2 m q \mathcal{E} -\sqrt{\mathcal{J}^4+2 \mathcal{J}^2 m
 \left(-m+2 m q^2-2 q \mathcal{E} \right)+m^2 \left(m^2-4 m q 
 \mathcal{E} +4 \mathcal{E} ^2\right)}<0. \label{con4}
\end{equation}
For this case, oscillation number then becomes 
\begin{eqnarray}
\mathcal{N}=
\frac{2m\sqrt{1 - q^2}}{\pi}\frac{1}{\sqrt{-\tilde{R}_{-}}}\bigg[
\tilde{R}_{-} \mathbb{E}\left(\frac{\tilde{R}_{+}}{\tilde{R}_{-}}\right)
+(1+\tilde{R}_{+})\big[\mathbb{K}\left(\frac{\tilde{R}_{+}}{\tilde{R}_{-}}\right)
-(1+\tilde{R}_{-}) \Pi(-\tilde{R}_{+},\frac{\tilde{R}_{+}}{\tilde{R}_{-}})\big]\bigg]\ . \nonumber
\end{eqnarray}
Where the $\tilde{R}_\pm$ are roots of the polynomial 
\begin{equation}
h(z) = (\mathcal{E}- mq)^2 - m^2(1-q^2)z(1+z)+ (2\mathcal{E}m q - m^2 q^2 -\mathcal{J}^2)(1+z).
\end{equation}
It is obvious here that the condition (\ref{con4}) actually comes from 
$ \frac{\tilde{R}_{+}}{\tilde{R}_{+} - \tilde{R}_{-}}<1 $.

Similarly as before, we can expand the oscillation number in different limits.
For the `long' string limit, where $\mathcal{E}\gg \mathcal{J}$, we can write the oscillation
number and invert it to get the string energy. It would include corrections to (\ref{AdSE})
in the powers of $\mathcal{J}$. We do not present the detailed expressions here for brevity.
As an example let us write the pulsating string energy for `long' strings in this background
\begin{equation}
 \mathcal{E}= 2 \mathcal{N} + \tilde{\mathcal{C}_1}(q)\sqrt{m \mathcal{N}} +  
\tilde{\mathcal{C}_2}(q)m + \left[\tilde{\mathcal{C}_3}(q) + \tilde{\mathcal{C}_4}(q)\frac{\mathcal{J}^2}{m^2}\right] 
\frac{m^{\frac{3}{2}}}{\sqrt{\mathcal{N}}} + \mathcal{O}(\mathcal{N}^{-\frac{3}{2}}).    \label{AdSE2}
\end{equation}
Where all other terms are same as in the previous subsection and the first correction of order 
$\mathcal{J}^2$ is given by the power series in $q$ as,
\begin{equation}
 \tilde{\mathcal{C}_4}(q)= 0.29509 + 0.06742 q + 0.03688 q^2 + ...
\end{equation}
It is worth mentioning that at the $\mathcal{O}(\frac{1}{\sqrt{\mathcal{N}}})$ there are no 
corrections in higher powers of $\mathcal{J}$. They start to appear in the following order and
are negligible in the large $\mathcal{N}$ regime.
\subsection*{String solutions with $q=1$}
\begin{figure}[h]
      \centering
        \begin{subfigure}[b]{0.235\textwidth}
                \includegraphics[width=\textwidth]{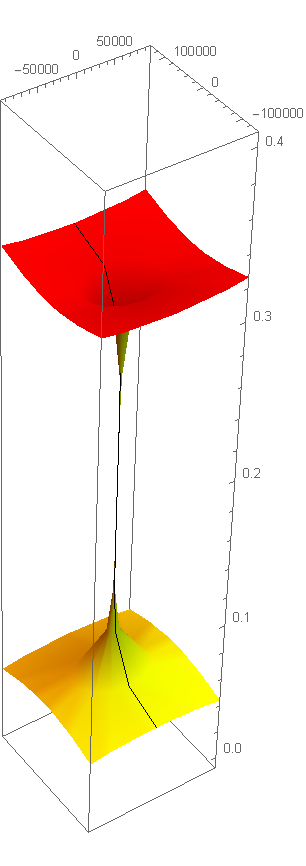}
                \caption{}
                \label{fig:4.3}
        \end{subfigure}%
         \begin{subfigure}[b]{0.22\textwidth}
                \includegraphics[width=\textwidth]{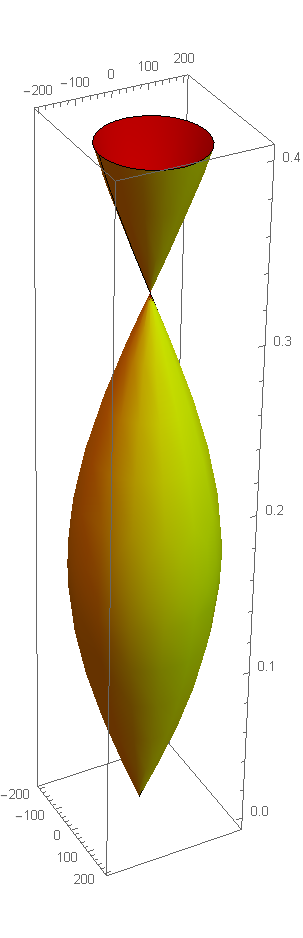}
                \caption{}
                \label{fig:4.4}
        \end{subfigure}
        \caption{Circular string solutions in $AdS_3$ with pure NS-NS flux as described by eq (\ref{pure2})
        and (\ref{pure3}).
        Both solutions have $m=1$ and $\mathcal{E}=2\times 10^3$. 
        While a) has $\mathcal{J}=0$ and does not show desired behaviour,
        b) has $\mathcal{J}= 64$ and re-gains pulsating motion due to the inclusion
        of angular momenta, however the pulsating strings are significantly smaller in size.}\label{fig:4}
\end{figure}

Let us start with the pulsating string in $AdS_3$ supported by pure NS-NS flux.
The equation of motion for $\rho$ takes the following form in this case 
\begin{equation}
 \dot{\rho}^2 - \frac{(\mathcal{E}  - m )^2}
{\cosh^2 \rho} + 2 \mathcal{E} m -m^2  = 0.
\end{equation}
This combined with $x = \sinh\rho$ leads to the equation of motion
\begin{equation}
 \dot{x}^2= \mathcal{E}^2+ (2\mathcal{E}m - m^2)x^2. \label{pure2}
\end{equation}
The solution now depends on the value of $d^2= (2\mathcal{E}m - m^2)$. For $d^2>0$ 
(as for example with $\mathcal{E}\gg m$ or, the long string limit) it can be written using the
initial condition $x(0)=0$,
\begin{eqnarray}
 \sinh\rho(\tau) =\pm \left[ \frac{e^{d \tau}}{2 d^2}-\frac{\mathcal{E}^2}{2}e^{- d \tau}\right]
\end{eqnarray}
which does not give rise to pulsation. On the other hand $d^2<0$
clearly gives an oscillatory solution of the form
\begin{equation}
 \sinh\rho(\tau) =\pm \frac{\mathcal{E}}{\tilde{d}}\sin{\tilde{d}\tau},~~~~~\tilde{d}=i d.
\end{equation}
  We can note here that if we put $q=1$ in (\ref{cond1}), the condition reduces to $(2\mathcal{E}-m) <0$,
which is the same constraint on pulsating motion we have just found. Now, our problem is that a string with large 
$\mathcal{E}$ (long) and small $m$ will not show pulsating behaviour. This problem can be solved, for example, by 
making large number of windings of the string. So for pure NS-NS background, only the large winding strings with high 
energy can become a `long' string .
In a related case, the string solution in $AdS_3\times S^1$ with $q=1$ solves the 
equation 
\begin{equation}
 \dot{x}^2= \mathcal{E}^2+ (2\mathcal{E}m - m^2-\mathcal{J}^2)x^2. \label{pure3}
\end{equation}
For the long string, inclusion of a large angular momentum will again lead to an
oscillatory pulsating solution. The trade-off however  is, with larger angular momentum attached to the
string, it will try to collapse onto itself, thereby reducing the effective size of
the string. However, the NS flux will continue trying to expand the string \cite{Maldacena:2000hw},
still resulting in finitely smaller size. This idea is illustrated in figure (\ref{fig:4}), 
but it  requires a deeper  understanding to be commented on conclusively.
The oscillation number for the small strings in $AdS_3$ 
reads
\begin{equation}
 \mathcal{N} = \frac{2}{\pi}\int^{r}_0 \frac{d x}{1+x^2}\sqrt{\mathcal{E}^2+ (2\mathcal{E}m - m^2)x^2}.
\end{equation}
Here $r=\sqrt{\frac{\mathcal{E}^2}{m^2- 2\mathcal{E}m}}$ is only real for small energies.
With a change of variable $x \rightarrow r x$, we can write this as
\begin{equation}
 \mathcal{N}= \frac{2}{\pi}r^2 d \int_{0}^{1}\frac{d x}{1+r^2 x^2}\sqrt{1-x^2} 
 = -\mathcal{E}+m -\sqrt{m^2- 2\mathcal{E}m}.
\end{equation}
Which has a small $\mathcal{E}$ expansion
\begin{equation}
 \mathcal{N} = \frac{\mathcal{E}^2}{2 m}+ \frac{\mathcal{E}^3}{2 m^2}+ \frac{5\mathcal{E}^4}{8 m^3}
 +\mathcal{O}(\mathcal{E}^5).
\end{equation}
And gives the exact relation for short string energy
\begin{equation}
 \mathcal{E}= \sqrt{2 m \mathcal{N}}-\mathcal{N}.
\end{equation}
As expected, the oscillation number for long strings becomes imaginary in this case.
\section{Conclusion}
In this note, we have discussed pulsating string solutions in subsectors of $AdS_3\times S^3$
with mixed flux and presented their dispersion relations. As expected, we could show the usual
dispersion relation in terms of the oscillation number and other conserved quantities receives
corrections due to the NS flux in the background. We have found out these corrections upto a
certain order in the the parameter governing the strength of the NS-NS flux
for both large and short string behaviour of the pulsating
string. For a complete 
understanding  into these string solutions, one would look for extract
the dual gauge theory operators for such strings in mixed flux background. Unfortunately, the dual gauge theory of
such a string background is not properly understood yet. The circular pulsating string solutions
in $AdS_5$ and the anomalous dimensions calculated from it played a significant part in
understanding relevant sectors of the dual SYM theory. We hope our solutions can be useful
in that regard also.

The straightforward extension of our work would be to find $(p,q)$ string solutions in this mixed
flux background which will be affected by both RR and NS-NS fluxes. This probe brane dynamics
should be interesting. The other way would be to verify our solutions by properly reducing the
oscillating string sigma model to that of a Neumann-Rosochatius integrable system 
\cite{Hernandez:2014eta,Hernandez:2015nba} using a pulsating type ansatz.
Also, the $q=1$ pulsating solutions should be
obtainable from a WZW model perspective. We hope to report on some of these issues in 
near future.

\section*{Acknowledgements}
It is a pleasure to acknowledge Suman Chatterjee, S. Pratik Khastgir and Abhishake Sadhukhan
for useful discussions. 
 \renewcommand{\theequation}{A-\arabic{equation}}
  \setcounter{equation}{0}  
  \section*{APPENDIX}  
\section*{Hamiltonian dynamics and pulsating strings}
We want to observe
the phase space of the pulsating string moving on the $AdS_3$ background
supported by NS-NS flux. We will sketch an outline of this discussion here
but will not talk about details.
The lagrangian of such a pulsating string would be given by
\begin{equation}
 \mathcal{L}=-\frac{1}{2}\left[- \cosh^2\rho \dot{t}^2 +
\dot\rho^2 + m^2\sinh^2\rho - 2 \dot{t} m q \sinh^2 \rho\right],
\end{equation}
\begin{figure}[h]
      \centering
        \begin{subfigure}[b]{0.4\textwidth}
                \includegraphics[width=\textwidth]{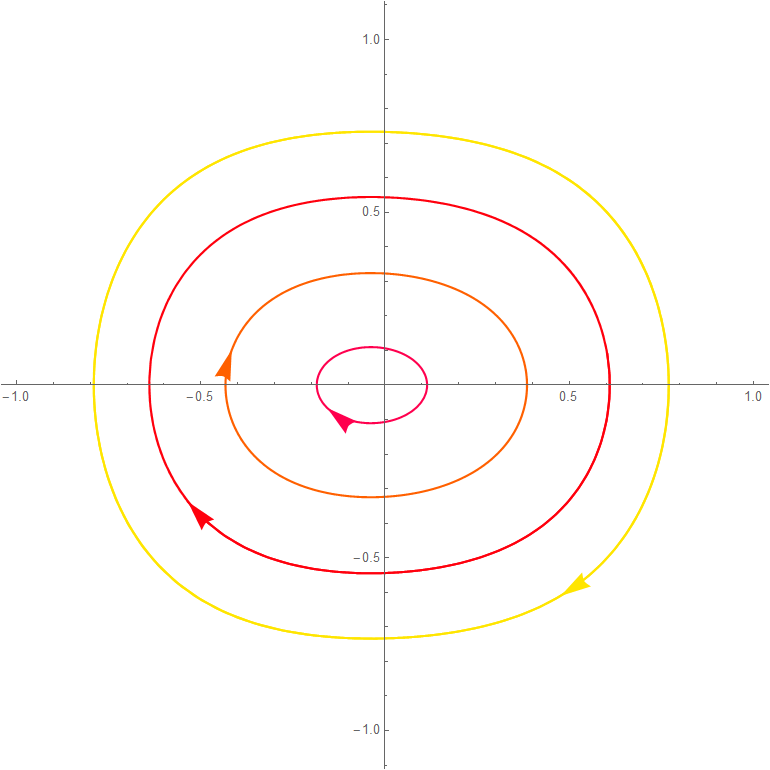}
                \caption{}
                \label{fig:6.3}
        \end{subfigure}~~~~
         \begin{subfigure}[b]{0.4\textwidth}
                \includegraphics[width=\textwidth]{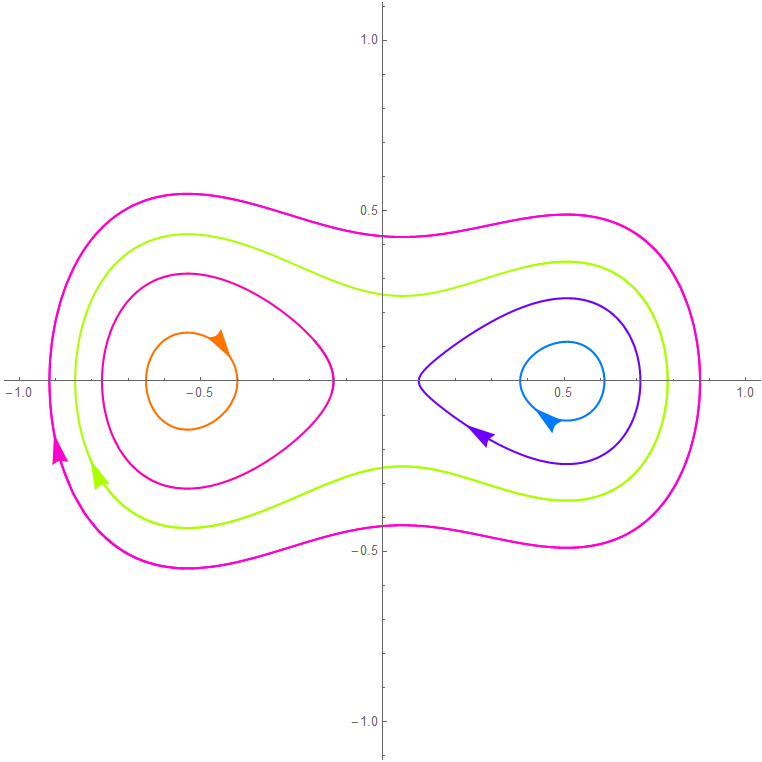}
                \caption{}
                \label{fig:6.4}
        \end{subfigure} 
        \\
        \begin{subfigure}[b]{0.35\textwidth}
                \includegraphics[width=\textwidth]{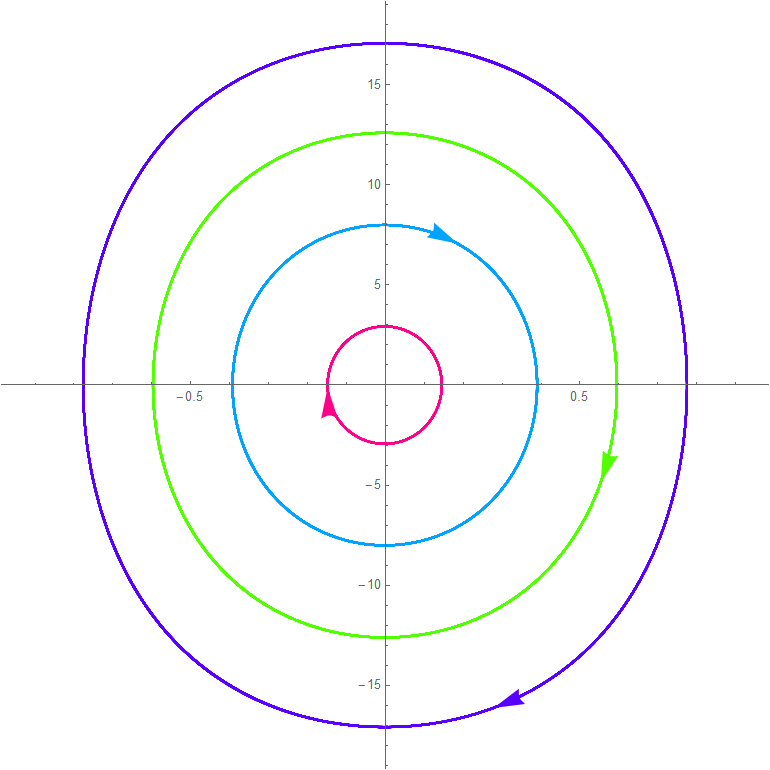}
                \caption{}
                \label{fig:6.5}
        \end{subfigure} ~~~~~
        \begin{subfigure}[b]{0.35\textwidth}
                \includegraphics[width=\textwidth]{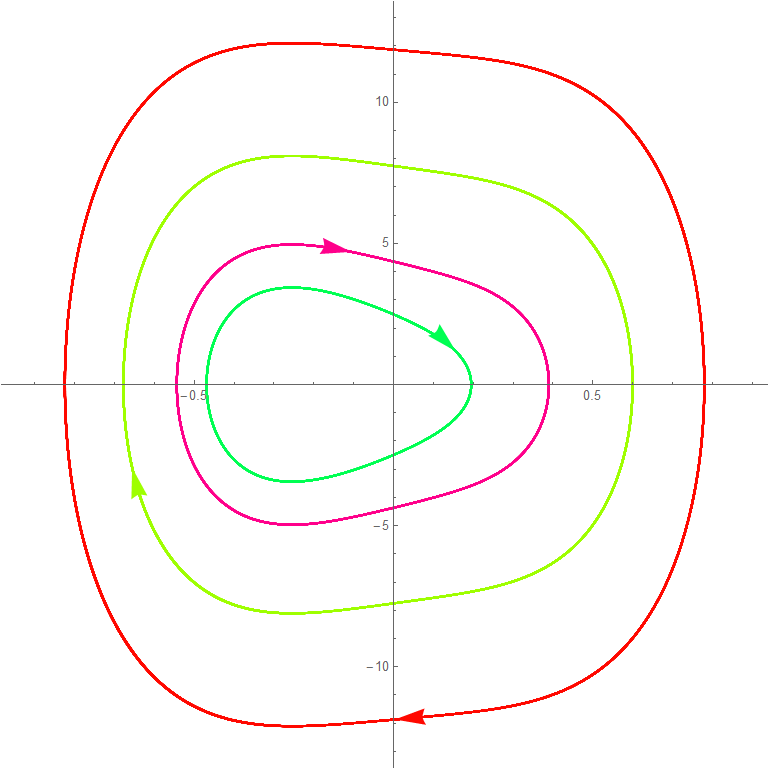}
                \caption{}
                \label{fig:6.6}
        \end{subfigure}
        \caption{Phase portraits ($P_\rho - \rho$) for the short string from (\ref{H}). a) is characterized by
        $\mathcal{E}=0.4$, $m=1$ and $q=0.05$ and shows nearly harmonic oscillator
        behaviour. As we increase the energy to $0.75$ in b) the classical symmetry 
        breaks and the additional equilibria occur. The original equilibrium is regained in 
        c) as we take the winding number to $m=20$. This equilibrium is maintained throughout
        even as we make the $q=0.999$ in d) as only the expansive nature of the NS-NS flux
        is not enough to disturb it.}\label{fig:5}
\end{figure}
\begin{figure}[h]
      \centering
        \begin{subfigure}[b]{0.35\textwidth}
                \includegraphics[width=\textwidth]{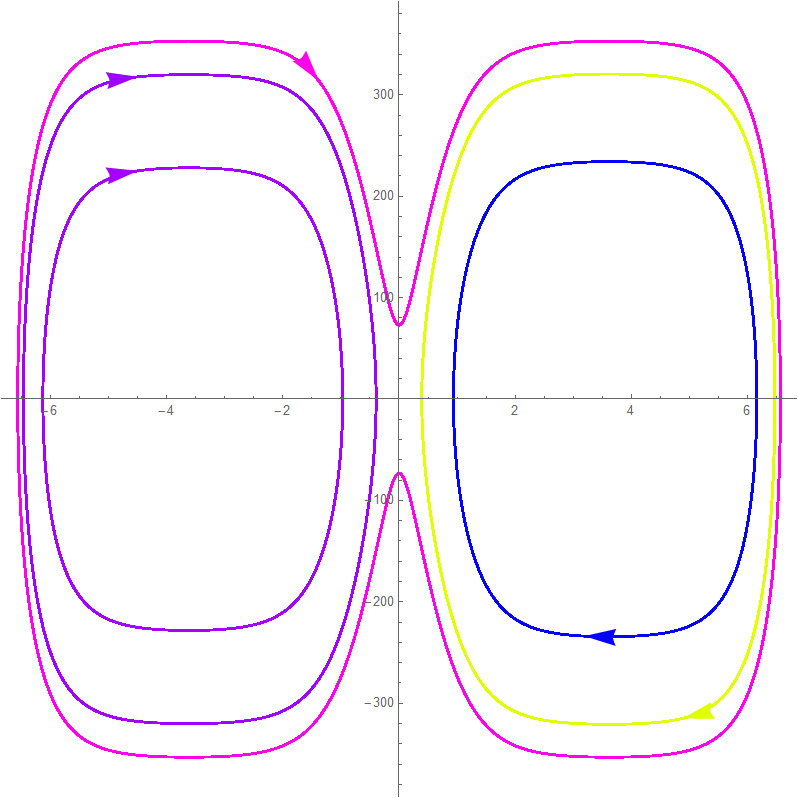}
                \caption{}
                \label{fig:7.3}
        \end{subfigure} ~~~~~~
         \begin{subfigure}[b]{0.35\textwidth}
                \includegraphics[width=\textwidth]{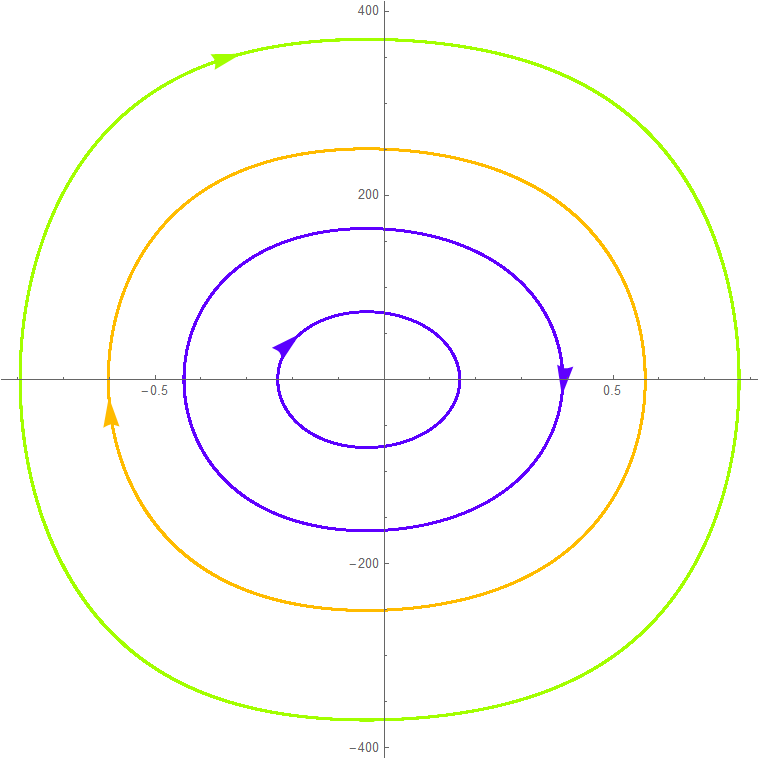}
                \caption{}
                \label{fig:7.4}
        \end{subfigure} 
        \\
        \begin{subfigure}[b]{0.35\textwidth}
                \includegraphics[width=\textwidth]{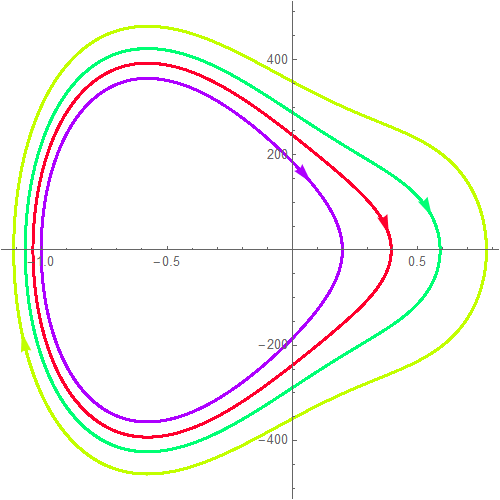}
                \caption{}
                \label{fig:8.4}
        \end{subfigure}
        \caption{Phase portraits ($P_\rho - \rho$) for the long string from (\ref{H}). a) is characterized by
        $\mathcal{E}= 200$, $m=1$ and $q=0.05$ and shows highly deformed  oscillator-like
        behaviour around the two minima. As we increase the winding number to $m=500$ in 
        b) the original equilibrium is regained, however in a highly distorted form.
        This equilibrium is maintained throughout even as we make the $q=0.999$ in c).}
        \label{fig:6}
\end{figure}
with proper choice of units. The corresponding hamiltonian can be written in terms of the conjugate
momenta as following,
\begin{equation}
 \mathcal{H}=\frac{1}{2}\Big[P_{\rho}^2+m^2\sinh^2\rho+\frac{(\mathcal{E}+mq\sinh^2\rho)}{\cosh^2\rho}
 (3\mathcal{E}-mq\sinh^2\rho)\Big]. \label{H}
\end{equation}
The hamiltonian equation of motion for $\rho$ can then be written as,
\begin{equation}
\ddot{\rho}= \frac{m q \mathcal{E}}{ \cosh\rho} - m^2 \cosh\rho \sinh\rho 
 + \frac{1}{\cosh^3\rho} \left[(m^2 q^2 + 3\mathcal{E}^2)\sinh\rho - 2m q \mathcal{E}\right].
\end{equation}
The phase portraits corresponding to the above hamiltonian has been given in figure (\ref{fig:5})
and (\ref{fig:6})
for various value of the parameters. We also have dscussed their behaviour qualitatively.
As a comparison, remember the  equation of motion for this pulsating string
could be cast into that of a non-linear oscillator
as we have already seen in section 3. The system has a form of
\begin{equation}
 \ddot{z}= \alpha z + \beta z^3.
\end{equation}
In general this equation is that of undamped Duffing type oscillator (for $\alpha < 0$) without any 
external forcing. With pulsating string ansatz, the coefficients $\alpha$ and $\beta$ turn out to 
be of useful form so that the solution can be written explicitly in terms of Jacobi elliptic functions
as we got in our detailed discussion.
Depending on the values of the parameters $m$, $q$ and the energy $\mathcal{E}$, the nature of the 
potential term changes, and so does the phase portraits when we talk in terms of point particles.
For sufficiently small $\beta$ the potential remains in the linear response regime and gives 
slightly deformed harmonic oscillator phase portraits.

Now looking at the potential, we can always say that $z= 0$ is a equilibrium point, but when
$\alpha\beta < 0$, the classical symmetry of the system is broken and two stable equilibria
occur at $z = \pm \sqrt{-\frac{\alpha}{\beta}}$. The stability analysis of these equilibria
can be done using the usual eigenvalue method. From a string point of view we can see that as 
the pulsating string attains more and more energy, the motion prefers the new minima of the 
potential. However if we increase the winding number suitably, the original equilibrium is 
regained as the state becomes `heavy'. If we increase the NS-NS flux, it tries to expand the
string, but is prohibited by the large tension. A remarkable fact here is that there is no 
forcing term, which guarantees that the chaotic attractor type situation does not occur here
for any value of the parameters.


\begin{thebibliography}{99} 
\bibitem{Maldacena:1997re}
  J.~M.~Maldacena,
  ``The Large N limit of superconformal field theories and supergravity,''
  Int.\ J.\ Theor.\ Phys.\  {\bf 38}, 1113 (1999)
  [Adv.\ Theor.\ Math.\ Phys.\  {\bf 2}, 231 (1998)]
  [hep-th/9711200].

\bibitem{Minahan:2002ve}
  J.~A.~Minahan and K.~Zarembo,
  ``The Bethe ansatz for N=4 superYang-Mills,''
  JHEP {\bf 0303}, 013 (2003)
  [hep-th/0212208].

\bibitem{Beisert:2003yb}
  N.~Beisert and M.~Staudacher,
  ``The N=4 SYM integrable super spin chain,''
  Nucl.\ Phys.\ B {\bf 670}, 439 (2003)
  [hep-th/0307042]. 
  
\bibitem{Bena:2003wd}
  I.~Bena, J.~Polchinski and R.~Roiban,
  ``Hidden symmetries of the AdS(5) x S**5 superstring,''
  Phys.\ Rev.\ D {\bf 69}, 046002 (2004)
  [hep-th/0305116].

\bibitem{Kazakov:2004qf}
  V.~A.~Kazakov, A.~Marshakov, J.~A.~Minahan and K.~Zarembo,
  ``Classical/quantum integrability in AdS/CFT,''
  JHEP {\bf 0405}, 024 (2004)
  [hep-th/0402207].

\bibitem{Zarembo:2004hp}
  K.~Zarembo,
  ``Semiclassical Bethe Ansatz and AdS/CFT,''
  Comptes Rendus Physique {\bf 5}, 1081 (2004)
  [Fortsch.\ Phys.\  {\bf 53}, 647 (2005)]
  [hep-th/0411191].

\bibitem{Beisert:2004hm}
  N.~Beisert, V.~Dippel and M.~Staudacher,
  ``A Novel long range spin chain and planar N=4 super Yang-Mills,''
  JHEP {\bf 0407}, 075 (2004)
  [hep-th/0405001].

\bibitem{Arutyunov:2004vx}
  G.~Arutyunov, S.~Frolov and M.~Staudacher,
  ``Bethe ansatz for quantum strings,''
  JHEP {\bf 0410}, 016 (2004)
  [hep-th/0406256].

\bibitem{Staudacher:2004tk}
  M.~Staudacher,
  ``The Factorized S-matrix of CFT/AdS,''
  JHEP {\bf 0505}, 054 (2005)
  [hep-th/0412188].

\bibitem{Beisert:2005fw}
  N.~Beisert and M.~Staudacher,
  ``Long-range psu(2,2|4) Bethe Ansatze for gauge theory and strings,''
  Nucl.\ Phys.\ B {\bf 727}, 1 (2005)
  [hep-th/0504190].

\bibitem{Beisert:2005tm}
  N.~Beisert,
  ``The SU(2|2) dynamic S-matrix,''
  Adv.\ Theor.\ Math.\ Phys.\  {\bf 12}, 945 (2008)
  [hep-th/0511082].

\bibitem{Hofman:2006xt}
  D.~M.~Hofman and J.~M.~Maldacena,
``Giant magnons,''
  J.\ Phys.\ A  {\bf 39}, 13095 (2006)
  [arXiv:hep-th/0604135].
  
  
\bibitem{Kruczenski:2004wg}
  M.~Kruczenski,
  \emph{``Spiky strings and single trace operators in gauge
  theories,''}
  JHEP {\bf 0508}, 014 (2005)
  [arXiv:hep-th/0410226].
  
\bibitem{Gubser:2002tv}
  S.~S.~Gubser, I.~R.~Klebanov and A.~M.~Polyakov,
  Nucl.\ Phys.\ B {\bf 636}, 99 (2002)
  [hep-th/0204051].
  
\bibitem{Minahan:2002rc}
  J.~A.~Minahan,
``Circular semiclassical string solutions on AdS(5) x S**5,''
  Nucl.\ Phys.\  B {\bf 648}, 203 (2003)
  [arXiv:hep-th/0209047]. 
  
\bibitem{Khan:2005fc}
  A.~Khan and A.~L.~Larsen,
  ``Improved stability for pulsating multi-spin string solitons,''
  Int.\ J.\ Mod.\ Phys.\ A {\bf 21}, 133 (2006)
  [hep-th/0502063].
  


\bibitem{Khan:2003sm}
  A.~Khan, A.~L.~Larsen,
  ``Spinning pulsating string solitons in AdS(5) x S**5,''
  Phys.\ Rev.\  {\bf D69}, 026001 (2004).
  [hep-th/0310019].

\bibitem{Engquist:2003rn}
  J.~Engquist, J.~A.~Minahan and K.~Zarembo,
``Yang-Mills duals for semiclassical strings on AdS(5) x S**5,''
  JHEP {\bf 0311}, 063 (2003)
  [arXiv:hep-th/0310188].

\bibitem{Arutyunov:2003za}
  G.~Arutyunov, J.~Russo, A.~A.~Tseytlin,
``Spinning strings in AdS(5) x S**5: New integrable system
relations,''
  Phys.\ Rev.\  {\bf D69}, 086009 (2004).
  [hep-th/0311004].

\bibitem{Dimov:2004xi}
  H.~Dimov and R.~C.~Rashkov,
``Generalized pulsating strings,''
  JHEP {\bf 0405}, 068 (2004)
  [arXiv:hep-th/0404012].

\bibitem{Smedback:1998yn}
  M.~Smedback,
  ``Pulsating strings on AdS(5) x S**5,''
  JHEP {\bf 0407}, 004 (2004)
  [arXiv:hep-th/0405102].
 
  
\bibitem{Kruczenski:2004cn}
  M.~Kruczenski and A.~A.~Tseytlin,
``Semiclassical relativistic strings in S**5 and long coherent
operators  in N = 4 SYM theory,''
  JHEP {\bf 0409}, 038 (2004)
  [arXiv:hep-th/0406189].
  
\bibitem{Chen:2008qq}
  B.~Chen and J.~B.~Wu,
  ``Semi-classical strings in $AdS_4 \times CP^3$,''
  JHEP {\bf 0809}, 096 (2008)
  [arXiv:0807.0802 [hep-th]].

\bibitem{Dimov:2009rd}
  H.~Dimov and R.~C.~Rashkov,
  ``On the pulsating strings in $AdS_4 x CP3$,''
  Adv.\ High Energy Phys.\  {\bf 2009}, 953987 (2009)
  [arXiv:0908.2218 [hep-th]].
  
\bibitem{Bobev:2004id}
  N.~P.~Bobev, H.~Dimov and R.~C.~Rashkov,
``Pulsating strings in warped $AdS(6) \times S^4$ geometry,''
  arXiv:hep-th/0410262. 
  
\bibitem{Arnaudov:2010by}
  D.~Arnaudov, H.~Dimov and R.~C.~Rashkov,
  ``On the pulsating strings in $AdS_5 x T^{1,1}$,''
  J.\ Phys.\ A {\bf 44}, 495401 (2011)
  [arXiv:1006.1539 [hep-th]].


\bibitem{Park:2005kt}
  I.~Y.~Park, A.~Tirziu and A.~A.~Tseytlin,
  ``Semiclassical circular strings in AdS(5) and 'long' gauge field strength operators,''
  Phys.\ Rev.\ D {\bf 71}, 126008 (2005)
  [hep-th/0505130].

\bibitem{Pradhan:2013sja}
  P.~M.~Pradhan and K.~L.~Panigrahi,
  ``Pulsating Strings With Angular Momenta,''
  Phys.\ Rev.\ D {\bf 88}, 086005 (2013)
  [arXiv:1306.0457 [hep-th]].

\bibitem{Beccaria:2010zn}
  M.~Beccaria, G.~V.~Dunne, G.~Macorini, A.~Tirziu and A.~A.~Tseytlin,
  ``Exact computation of one-loop correction to energy of pulsating strings in $AdS_5 x S^5$,''
  J.\ Phys.\ A {\bf 44}, 015404 (2011)
  [arXiv:1009.2318 [hep-th]].

 
\bibitem{Giardino:2011jy}
  S.~Giardino and V.~O.~Rivelles,
  ``Pulsating Strings in Lunin-Maldacena Backgrounds,''
  JHEP {\bf 1107}, 057 (2011)
  [arXiv:1105.1353 [hep-th]].


\bibitem{Banerjee:2014bca}
  A.~Banerjee and K.~L.~Panigrahi,
  ``On the rotating and oscillating strings in (AdS$_{3}$  x S$^{3}$)$_{\kappa}$,''
  JHEP {\bf 1409}, 048 (2014)
  [arXiv:1406.3642 [hep-th]].

\bibitem{Panigrahi:2014sia} 
  K.~L.~Panigrahi, P.~M.~Pradhan and M.~Samal,
  ``Pulsating strings on (AdS$_{3}$ × S$^{3}$)$_{ϰ}$,''
  JHEP {\bf 1503}, 010 (2015)
  [arXiv:1412.6936 [hep-th]].

\bibitem{Banerjee:2014rza} 
  A.~Banerjee, S.~Biswas and K.~L.~Panigrahi,
  ``Semiclassical Strings in Supergravity PFT,''
  Eur.\ Phys.\ J.\ C {\bf 74}, no. 10, 3115 (2014)
  [arXiv:1403.7358 [hep-th]].
  
\bibitem{Maldacena:2000hw}
J.~M. Maldacena and H.~Ooguri, {\it {Strings in {$AdS_3$} and {$SL(2,R)$} WZW
  model 1.: The Spectrum}},  {\em J.Math.Phys.} {\bf 42} (2001) 2929--2960,
  [\href{http://xxx.lanl.gov/abs/hep-th/0001053}{{\tt hep-th/0001053}}].
  
\bibitem{misc6}
  B.~-H.~Lee, R.~R.~Nayak, K.~L.~Panigrahi, C.~Park,
  \emph{``On the giant magnon and spike solutions for strings on AdS(3) x
  S**3,''}
  JHEP {\bf 0806}, 065 (2008).
  [arXiv:0804.2923 [hep-th]]. 
  
  \bibitem{David:2008yk}
J.~R. David and B.~Sahoo, {\it {Giant magnons in the D1-D5 system}},  {\em
  JHEP} {\bf 0807} (2008) 033, [\href{http://xxx.lanl.gov/abs/0804.3267}{{\tt
  arXiv:0804.3267}}]

\bibitem{David:2010yg}
J.~R. David and B.~Sahoo, {\it {S-matrix for magnons in the D1-D5 system}},
  {\em JHEP} {\bf 1010} (2010) 112,
  [\href{http://xxx.lanl.gov/abs/1005.0501}{{\tt arXiv:1005.0501}}].

\bibitem{Abbott:2012dd}
M.~C. Abbott, {\it {Comment on Strings in {$AdS_3\times S_3 \times S_3 \times
  S_1$} at One Loop}},  {\em JHEP} {\bf 1302} (2013) 102,
  [\href{http://xxx.lanl.gov/abs/1211.5587}{{\tt arXiv:1211.5587}}].

\bibitem{Beccaria:2012kb}
M.~Beccaria, F.~Levkovich-Maslyuk, G.~Macorini, and A.~Tseytlin, {\it {Quantum
  corrections to spinning superstrings in {$AdS_3 \times S^3 \times M^4$}:
  determining the dressing phase}},  {\em JHEP} {\bf 1304} (2013) 006,
  [\href{http://xxx.lanl.gov/abs/1211.6090}{{\tt arXiv:1211.6090}}].

\bibitem{Beccaria:2012pm}
M.~Beccaria and G.~Macorini, {\it {Quantum corrections to short folded
  superstring in {$AdS_3 \times S^3 \times M^4$} }},  {\em JHEP} {\bf 1303}
  (2013) 040, [\href{http://xxx.lanl.gov/abs/1212.5672}{{\tt
  arXiv:1212.5672}}].
  
\bibitem{Abbott:2013ixa}
M.~C. Abbott, {\it {The {$AdS_{3} \times S^{3}\times S^{3}\times S^{1}$}
  Hern\'{a}ndez-L\'{o}pez phases: a semiclassical derivation}},  {\em J. Phys.}
  {\bf A46} (2013) 445401, [\href{http://xxx.lanl.gov/abs/1306.5106}{{\tt
  arXiv:1306.5106}}].

\bibitem{Rughoonauth:2012qd}
N.~Rughoonauth, P.~Sundin, and L.~Wulff, {\it {Near BMN dynamics of the AdS(3)
  x S(3) x S(3) x S(1) superstring}},  {\em JHEP} {\bf 1207} (2012) 159,
  [\href{http://xxx.lanl.gov/abs/1204.4742}{{\tt arXiv:1204.4742}}].

\bibitem{Cardona:2014gqa} 
  C.~Cardona,
  ``Pulsating strings from two dimensional CFT on $(T^4)^N/S(N)$,''
  Nucl.\ Phys.\ B {\bf 893}, 512 (2015)
  [arXiv:1408.5035 [hep-th]].
 
\bibitem{Sfondrini:2014via} 
  A.~Sfondrini,
  ``Towards integrability for ${\rm Ad}{{{\rm S}}_{{\bf 3}}}/{\rm CF}{{{\rm T}}_{{\bf 2}}}$,''
  J.\ Phys.\ A {\bf 48}, no. 2, 023001 (2015)
  [arXiv:1406.2971 [hep-th]].
 
 \bibitem{Cagnazzo:2012se}
A.~Cagnazzo and K.~Zarembo, {\it {B-field in {$AdS_3/CFT_2$} Correspondence and
  Integrability}},  {\em JHEP} {\bf 1211} (2012) 133,
  [\href{http://xxx.lanl.gov/abs/1209.4049}{{\tt arXiv:1209.4049}}].

\bibitem{Wulff:2014kja}
L.~Wulff, {\it {Superisometries and integrability of superstrings}},
  \href{http://xxx.lanl.gov/abs/1402.3122}{{\tt arXiv:1402.3122}}.

  \bibitem{Hoare:2013pma}
B.~Hoare and A.~Tseytlin, {\it {On string theory on {$AdS_3 \times S^3\times
  T^4$} with mixed 3-form flux: Tree-level S-matrix}},  {\em Nucl.Phys.} {\bf
  B873} (2013) 682--727, [\href{http://xxx.lanl.gov/abs/1303.1037}{{\tt
  arXiv:1303.1037}}].

\bibitem{Hoare:2013ida}
B.~Hoare and A.~Tseytlin, {\it {Massive S-matrix of $AdS_3 \times S^3 \times
  T^4$ superstring theory with mixed 3-form flux}},  {\em Nucl.Phys.} {\bf
  B873} (2013) 395--418, [\href{http://xxx.lanl.gov/abs/1304.4099}{{\tt
  arXiv:1304.4099}}].

\bibitem{Bianchi:2014rfa}
L.~Bianchi and B.~Hoare, {\it {$AdS_3 \times S^3 \times M^4$ string S-matrices
  from unitarity cuts}},  \href{http://xxx.lanl.gov/abs/1405.7947}{{\tt
  arXiv:1405.7947}}.

\bibitem{Hoare:2013lja}
B.~Hoare, A.~Stepanchuk, and A.~Tseytlin, {\it {Giant magnon solution and
  dispersion relation in string theory in {$AdS_3\times S^3\times T^4$} with
  mixed flux}},  {\em Nucl.Phys.} {\bf B879} (2014) 318--347,
  [\href{http://xxx.lanl.gov/abs/1311.1794}{{\tt arXiv:1311.1794}}].

\bibitem{Babichenko:2014yaa}
A.~Babichenko, A.~Dekel, and O.~Ohlsson~Sax, {\it {Finite-gap equations for
  strings on $AdS_3 x S^3 x T^4$ with mixed 3-form flux}},
  \href{http://xxx.lanl.gov/abs/1405.6087}{{\tt arXiv:1405.6087}}.

\bibitem{Borsato:2014hja} 
  R.~Borsato, O.~Ohlsson Sax, A.~Sfondrini and B.~Stefanski,
  ``The complete AdS$_{3} \times$ S$^3 \times$ T$^4$ worldsheet S matrix,''
  JHEP {\bf 1410}, 66 (2014)
  [arXiv:1406.0453 [hep-th]].
  
\bibitem{Lloyd:2014bsa} 
  T.~Lloyd, O.~Ohlsson Sax, A.~Sfondrini and B.~Stefanski, Jr.,
  ``The complete worldsheet S matrix of superstrings on $AdS_3 x S^3 x T^4$ with mixed three-form flux,''
  Nucl.\ Phys.\ B {\bf 891}, 570 (2015)
  [arXiv:1410.0866 [hep-th]].
 
\bibitem{David:2014qta} 
  J.~R.~David and A.~Sadhukhan,
  ``Spinning strings and minimal surfaces in $AdS_3$ with mixed 3-form fluxes,''
  JHEP {\bf 1410}, 49 (2014)
  [arXiv:1405.2687 [hep-th]].
  
\bibitem{Ahn:2014tua} 
  C.~Ahn and P.~Bozhilov,
  ``String solutions in $AdS_3 x S^3 x T^4$ with NS-NS B-field,''
  Phys.\ Rev.\ D {\bf 90}, no. 6, 066010 (2014)
  [arXiv:1404.7644 [hep-th]].

\bibitem{misc9}
  A.~Banerjee, K.~L.~Panigrahi and P.~M.~Pradhan,
  ``Spiky strings on $AdS_3 \times S^3$ with NS-NS flux,''
  Phys.\ Rev.\ D {\bf 90}, no. 10, 106006 (2014)
  [arXiv:1405.5497 [hep-th]].

\bibitem{Hernandez:2014eta} 
  R.~Hernández and J.~M.~Nieto,
  ``Spinning strings in $AdS_3 \times S^3$ with NS–NS flux,''
  Nucl.\ Phys.\ B {\bf 888}, 236 (2014)
  [Nucl.\ Phys.\ B {\bf 895}, 303 (2015)]
  [arXiv:1407.7475 [hep-th]].
  
\bibitem{Hernandez:2015nba} 
  R.~Hernandez and J.~M.~Nieto,
  ``Elliptic solutions in the Neumann–Rosochatius system with mixed flux,''
  Phys.\ Rev.\ D {\bf 91}, no. 12, 126006 (2015)
  [arXiv:1502.05203 [hep-th]].



\end{thebibliography}
\end{document}